# Quantum oscillations from the reconstructed Fermi surface in electron-doped cuprate superconductors

J S Higgins[1], M K Chan[2], Tarapada Sarkar[1], R D McDonald[2], R L Greene[1], N P Butch[1,3]

[1] Center for Nanophysics and Advanced Materials, Department of Physics, University of Maryland, College Park, Maryland 20742 USA

[2] The National High Magnetic Field Laboratory, Los Alamos National Laboratory, Los Alamos, New Mexico 87545 USA

[3] NIST Center for Neutron Research, National Institute of Standards and Technology, Gaithersburg, Maryland 20899 USA

E-mail: Nicholas.butch@nist.gov

## Abstract

We have studied the electronic structure of electron-doped cuprate superconductors via measurements of high-field Shubnikov-de Haas oscillations in thin films.  In optimally doped $Pr_{2-x}Ce_xCuO_{4\pm\delta}$ and $La_{2-x}Ce_xCuO_{4\pm\delta}$, quantum oscillations indicate the presence of a small Fermi surface, demonstrating that electronic reconstruction is a general feature of the electron-doped cuprates, despite the location of the superconducting dome at very different doping levels. Negative high-field magnetoresistance is correlated with an anomalous low-temperature change in scattering that modifies the amplitude of quantum oscillations.  This behaviour is consistent with effects attributed to spin fluctuations.

## Introduction

The fundamentals of the electronic interactions in the copper oxide family of superconductors (cuprates) remain a subject of intense debate. One open question of particular importance is the nature of the normal state from which superconductivity emerges – whether it resembles a metal, having well-defined electron-like excitations, or a more exotic electronic phase. The general landscape is complicated by the presence of magnetic order, which can disrupt the stability of the metallic phase.  In hole-doped cuprates, the presence of a 'pseudogap' phase leads to even more uncertainty about the nature of the normal state. The ultimate question is what are the underlying building blocks of high temperature superconductivity?

The electron-doped cuprates $R_{2-x}Ce_xCuO_{4\pm\delta}$, where R is a light rare earth atom such as La, Nd, or Pr are a useful testbed for developing an understanding of the minimum ingredients behind cuprate superconductivity [1] since no pseudogap phase is found. A schematic superconducting phase diagram for the electron-doped cuprates is shown in Figure 1(a). Long-range antiferromagnetic (AFM) order is suppressed at the onset of superconductivity [2], although signatures of short range order appear to persist beyond optimal doping [3-5]. The value of $x$ for which the superconducting transition $T_c$ is maximum is $x_o$ = 0.11 for La, 0.145 for Nd, and 0.15 for Pr. The actual cerium concentrations that determine the superconducting dome vary among the electron-doped cuprates, although the general shape of the phase diagram is constant [6].

As cerium doping increases from the parent compound, long-range AFM order (AFM correlation length, $\xi_{AFM}$, greater than 400 lattice constants, a) gives way to a superconducting dome and finite short-range AFM correlations ($\xi_{AFM} < 50$ a) that persist up to a critical concentration ~ $x_c$, ending at a possible quantum critical point (QCP) where the Fermi surface (FS) reconstructs. Evidence for a Fermi surface reconstruction (FSR) and possible QCP in this portion of the phase diagram has been reported in measurements of the Hall effect [7-10], optical conductivity [4, 5], thermoelectric power [11], angular magnetoresistance [12, 13], and angle-resolved photoemission spectroscopy (ARPES) [14, 15]. At this doping ($x_c$), the sign of the Hall coefficient changes from negative to positive. The p-type regime ($x > x_c$) is characterized by a simple one-band response in the low temperature limit, whereas the n-type is multiband for $x_o < x < x_c$. This transport behaviour is consistent with the ARPES-derived Fermi surface diagrams, in which zone folding transforms one hole-like FS on the highly doped p-type side to multiple FSs on the lower doped n-type side. At concentrations beyond the superconducting dome, a normal metallic Fermi liquid ground state is observed that exhibits quantum critical scaling as a function of doping and magnetic field [16]. In contrast to the Fermi liquid metal found beyond the superconducting dome, at lower $x$ the unusual linear in T temperature dependence of the electrical resistivity implies that the normal state is not a conventional metal and that scattering is dominated by spin fluctuations [17].

One way to directly track the evolution of the FS is through Shubnikov-de Haas (SdH) oscillations in the magnetotransport. At sufficiently low temperatures, the motion of electrons in a magnetic field is quantized in to Landau levels. As the magnetic field increases, these levels cross the FS resulting in an oscillation in the density of states at the Fermi energy. The frequency of these oscillations relate to the area of the FS through the Onsager relation, $F = S\hbar/2\pi e$, where $F$ is the frequency of the oscillation and

$S$ is the extremal area of the FS. Helm *et al*. reported oscillations in interlayer magnetoresistance of superconducting single crystal NCCO, for dopings near $x_o$ ($x = 0.145$) to overdoped ($x > x_c$, $x = 0.17$) [18-20]. They found a predominantly low frequency (~300 T) oscillation near optimal doping ($x_0$), corresponding to ~ 1% of the Brillouin zone. Comparisons with ARPES data suggest that this FS is the hole-pocket in the reconstructed FS [inset of Fig. 1(a)]. At higher dopings, a high frequency (~ 11 kT) oscillation was observed, corresponding to ~ 41% of the Brillouin zone. Initially this was thought to be the unreconstructed FS at $x > x_c$, however, the low and high frequency oscillations were found to co-exist for all dopings measured. Indication of a FSR from SdH oscillations would be detected as a change in the oscillation frequency, from 300 T to 11 kT, where the material goes from n-type to p-type. Rather than measuring the unreconstructed FS, magnetic breakdown involving tunneling between the electron and hole pockets was suggested to be responsible for the high frequency oscillations [19]. Additionally, Breznay *et al*. observed SdH oscillations of the in-plane magnetoresistance in PCCO thin films ($x \approx x_o$, $x = 0.14$) [21]. However, only the small hole pockets are observable in quantum oscillations for both PCCO and NCCO at $x_0$. In contrast, for $x > x_c$, ARPES detects the unreconstructed FS (single hole-like pocket) [14], but no signs of it have been measured via quantum oscillations (i.e., a single high-frequency oscillation).

To understand how the choice of rare earth element affects the FS, we report here measurements of quantum oscillations in thin films of PCCO for $x < x_c$ and $x > x_c$ and in optimally doped LCCO, which can only be stabilized in film form [22, 23]. We find that at $x_o$ in both LCCO and PCCO, there is evidence for Shubnikov-de Haas oscillations with comparable frequencies to those seen in NCCO, firmly establishing that optimally electron-doped cuprate superconductivity arises from the same metallic normal state. For $x > x_c$, we observe no clear quantum oscillations. Near $x_c$ we observe an anomalous increase in the low-temperature scattering rate, consistent with the persistence of spin fluctuations that plays a role in the FSR.

## Method

Epitaxial thin films of *c*-axis oriented PCCO and LCCO were grown on (100) SrTiO$_3$ substrates using pulsed laser deposition and subsequent annealing, as described elsewhere [24]. Stoichiometric targets, with the desired nominal cerium concentration, were used and the effective doping was further determined through x-ray [24] and Hall effect [7] measurements. Film thicknesses were 150 – 200 nm, as determined by cross-sectional scanning electron microscopy. All films were characterized using x-ray, AC-susceptibility, resistivity, and Hall measurements. The films were patterned into Hall bar geometries.

High-field magnetoresistance (MR) measurements were performed at the Pulsed-Field Facility at Los Alamos National Laboratory. The field was applied along the $c$-axis of the samples and measurements of the in-plane MR were performed using both 65T and 80 T pulsed field magnets with a $^3$He refrigerator.

## Results and Discussion

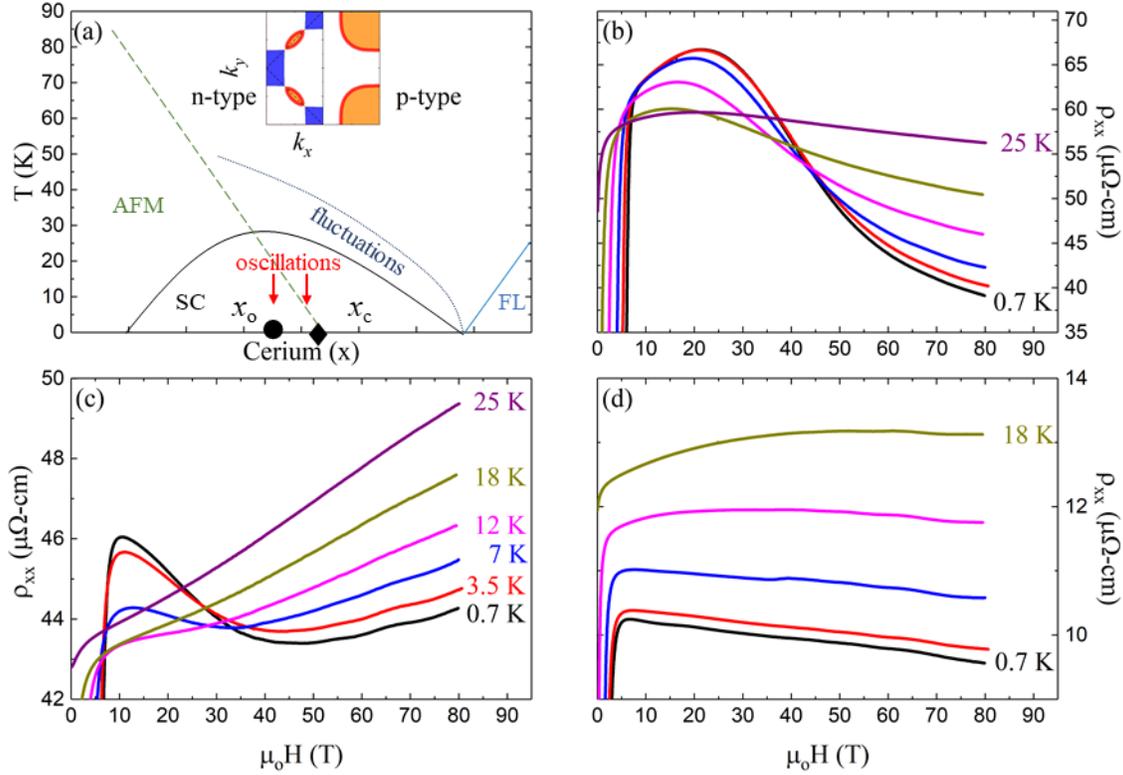

Figure 1: (a) Phase diagram for the electron-doped cuprates showing the antiferromagnetic insulator phase and the extent of short range magnetic correlations with respect to the superconducting dome. The arrows point to the dopings in which we see oscillations. Inset shows the Fermi surface of the first Brillouin zone with the left side representing the reconstructed, magnetically zone-folded surface, and the right side representing the unreconstructed surface. In-plane magnetoresistance of three electron-doped films in applied magnetic fields up to 80 T (B||$c$-axis): (b) LCCO, $x = 0.11$; (c) PCCO, $x = 0.15$; (d) PCCO, $x = 0.16$.

Figure 1(b-d) shows the measured MR up to 80 T for select concentrations. The optimally doped PCCO ($x = 0.15$) and the LCCO ($x = 0.11$) samples show a hump-like behaviour at low fields and temperatures. The PCCO sample also shows a minimum in the MR where it goes from negative to positive. In contrast,

the overdoped PCCO ($x = 0.16$) sample shows a negative MR up to highest measured fields at low temperatures. At higher dopings, the MR is positive. These MR features have been reported before [25-27]. In particular, linear negative MR in the non-superconducting n-type Hall effect regime has been connected to spin scattering, and associated with spin fluctuation scattering [26]. The hump in MR is not observed in the p-type concentrations.

Quantum oscillations from optimally-doped PCCO are similar to those reported from measurements on NCCO and undoped PCO (Fig. 2). In order to see the oscillations, a background is subtracted by fitting the MR data with $R(T, B) = P(T, B) \cdot (1 + A_{osc})$, where P(T, B) is approximated by a fifth order polynomial and $A_{osc}$ is the field B and temperature T dependent oscillatory component. $A_{osc}$ follows the standard Lifshitz-Kosevich (LK) formalism:

$$A_{osc} = \frac{\Delta R}{R_{bg}} = A_0 \cdot R_D \cdot R_T \cos(2\pi F/B - \gamma) \qquad (1)$$

Here $A_0$ is a sample-dependent constant, $R_D$ is the Dingle factor $[e^{-\left(\frac{\pi m^*}{e\tau_D} \cdot \frac{1}{B}\right)}]$ defining the field dependent envelope function, $R_T$ is the LK temperature factor $[\frac{2\pi^2 k_B m^* T}{\hbar eB}/sinh\left(\frac{2\pi^2 k_B m^* T}{\hbar eB}\right)]$, and $\gamma$ is the phase.

The measured oscillations are clear, and are on the order of $10^{-4}$ of the total resistance. For optimally doped PCCO, the observed oscillations have a low frequency of 294 T, corresponding to approximately 1% of the square Brillouin zone area. This represents a small FS that is consistent with the reconstructed hole pocket observed in ARPES in NCCO, and FS areas from QO measurements on NCCO [18], PCCO [21], and PCO [25]. The temperature dependence yields an effective mass of 1 electron mass, comparable to values from NCCO and PCO.

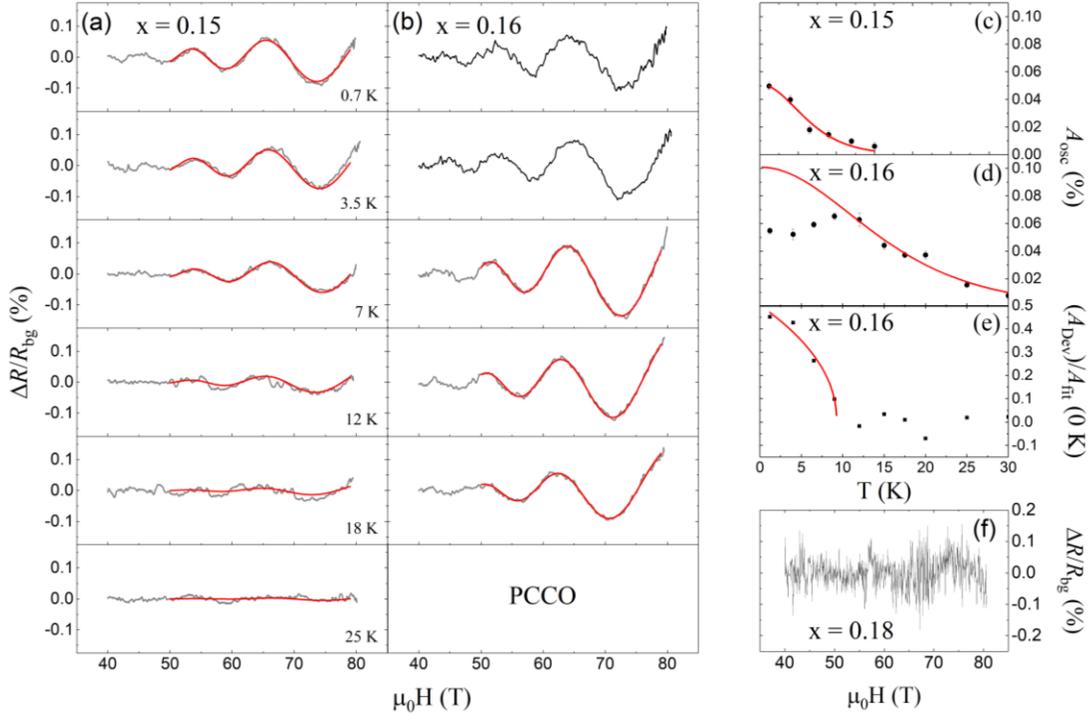

Figure 2: Oscillations in the MR after the background has been subtracted: (a) PCCO, $x = 0.15$; (b) PCCO, $x = 0.16$. Red lines are global fits to the LK formula for each doping. Panels (c) and (d) show the temperature dependence of the oscillation amplitudes. The red lines are fits to the temperature dependent factor ($R_T$) in equation 1. The fit is extended to low temperatures for the $x = 0.16$ sample (d) to emphasize the oscillation amplitude deviation. (e) Normalized reduction of the quantum oscillation amplitude, which may be proportional to a temperature-dependent volume fraction of sample having a large p-type Fermi surface. (f) Background subtracted MR of the overdoped PCCO $x = 0.18$ at $T = 0.7$ K. No clear oscillations are observed.

Next, we look at PCCO $x = 0.16$, which is slightly overdoped, but still shows multi-band behaviour. It is important to point out the FS reconstruction does not sit at optimal doping. Here clear quantum oscillations are also observed, having a slightly smaller frequency of 267 T, indicative of a smaller FS. In the temperature regime where the MR begins to crossover from negative to positive, the LK formalism holds, yielding an effective mass of 0.4 $m_e$. However, below 10 K where the sample shows a negative MR, the oscillation amplitude deviates from the high-temperature trend, and the phase shifts.

One possible explanation for the reduction in quantum oscillation amplitude at low temperature is that some volume fraction of the sample transforms to a phase with a different Fermi surface, as calculated in

Fig. 2(e). Being near $x_c$, this most likely entails conversion to the large p-type Fermi surface. Some phase heterogeneity is well known in electron doped cuprates, particularly small amounts of long-range antiferromagnetism in NCCO [3]. In the same way, small portions of nonmagnetic p-type phase may also be distributed throughout samples near $x_c$.

Another possibility is that the effective mobility decreases at low temperatures, which is readily interpreted as an increased or additional scattering rate. It is very conspicuous that this anomalous scattering only occurs at temperatures and magnetic fields where the MR is negative. Dagan *et al*. [26] have shown that the strong negative MR in the MR hump region is attributable to an isotropic spin scattering process. It has also been shown that a linear negative MR in the underdoped side of the phase diagram can be attributed to AFM correlations [27]. The close proximity of this sample to the FSR and the deviation of the oscillations from the LK formula in the negative MR regions together suggest that dynamic spin fluctuations, separate from static short range AFM correlations, are responsible for the oscillation amplitude suppression. In contrast, the optimally doped PCCO sample does not deviate from the LK formula since the oscillations are only observed where the MR is positive. One could therefore think of these low temperatures in the $x = 0.16$ PCCO sample as a crossover into a strong spin fluctuation regime, whereas at optimal doping the temperature scale is a lot higher, and therefore all oscillations occur inside this regime.

The change in oscillation frequency of the 0.16 sample relative to the optimally doped 0.15 sample is larger than that reported in NCCO [28], where the doping dependence is nearly flat at that concentration, but the trend is consistent when we consider that the superconducting dome shifts slightly with doping among the electron-doped cuprates, resulting in different dopings for $x_0$ and $x_c$. For $x > 0.16$ in NCCO, a significant decrease in the oscillation frequency is also reported.

In the far overdoped regime, for PCCO $x = 0.18$, the high-field MR is always positive for the temperatures measured (not shown). This highlights the peculiar role of the slightly overdoped PCCO $x = 0.16$, which is the only concentration measured with negative MR but no hump. Together with the extrapolated endpoint of short-range correlations [12] and the change in Hall sign [7], this suggests that $x = 0.16$ sits on top of the FS critical point. In analogy to NCCO, beyond $x = 0.16$, a large hole-like unreconstructed FS is expected. We did not detect any clear quantum oscillations in these films, even though electrical properties were similar to those at lower doping. Similarly, Helm *et al*. did not observer quantum oscillations due to the large FS in NCCO, and instead saw signatures of magnetic breakdown [19]. In our case, the mobility of our $x = 0.18$ sample may have been low enough that the Dingle factor in Eq. 1 would suppress the oscillations to within the noise of the measurements, not allowing any high frequency oscillations from either a large FS or from magnetic breakdown.

To assess the universality of this behaviour, we also studied quantum oscillations in LCCO. In Fig. 3, the quantum oscillations of optimally-doped LCCO are shown. Fits to the data yield a frequency of 305 T and effective mass of 0.8 $m_e$. These values are similar to those of optimally-doped PCCO, underscoring the similarity of the electronic structure responsible for the superconducting state. The FSR in LCCO is estimated to be near $x = 0.14$ [8]. On the other hand, a dramatic difference is that the temperature dependence of the amplitude deviates from LK below 7 K. In this regard, the optimally doped LCCO is similar to the $x = 0.16$ PCCO in that both exhibit negative MR up to high fields, whereas the optimally doped PCCO does not. Again, the most likely source of mobility reduction is increased scattering due to spin fluctuations. It is important to note that the high-field behaviour of LCCO films is largely unexplored, and the extent of the negative MR behaviour is a topic of ongoing research.

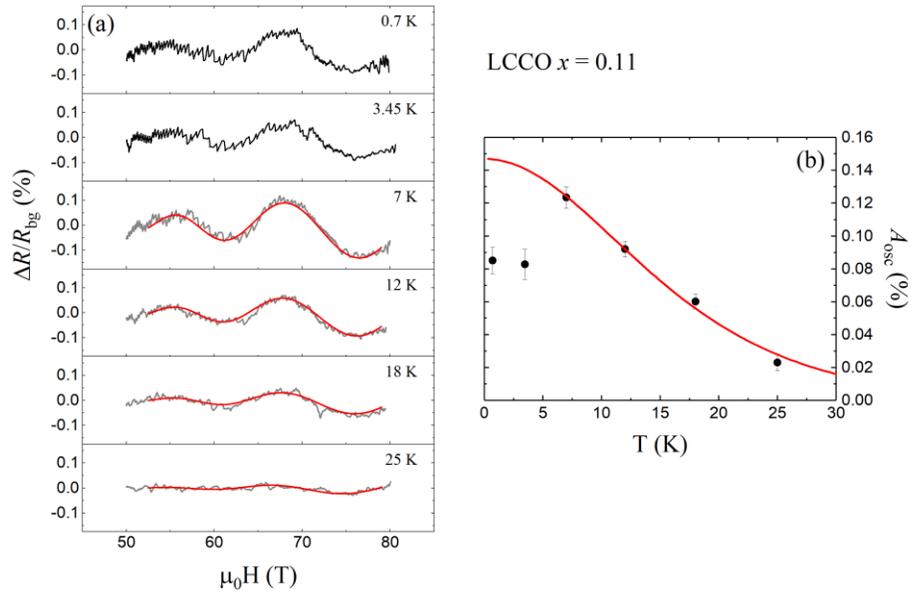

Figure 3: (a) Oscillations in the MR after the background has been subtracted for the LCCO, $x = 0.11$, sample. Red lines are global fits to the LK formula. (b) Temperature dependence of the oscillation amplitudes. The red lines are fits to the temperature dependent factor ($R_T$) in equation 1. The fit is extended to low temperatures to emphasize the oscillation amplitude deviation.

It is instructive to compare our results to prior quantum oscillations measurements. A comprehensive set of measurements was performed on bulk crystals of NCCO by Helm *et al*. [18]. The FS area and effective mass that we determine for optimally doped PCCO and LCCO correspond to the values measured in optimally doped NCCO. In addition, measurements on overdoped PCCO $x = 0.18$ show no clear evidence for quantum oscillations, as seen in overdoped NCCO. Both pieces of evidence strongly support a notion of a universal electronic structure in the electron doped cuprates. Our FS area values are also comparable to those seen in PCO films [25]. Compared to recently published results on PCCO films of $x = 0.14$ [21], where the frequency is 255T and $m_e = 0.43$, our PCCO $x = 0.15$ data disagree substantially. However, note that near optimal doping, a hump is typically observed in the MR, and the absence of such a feature suggests that the sample in Ref. 9 actually may have a doping level comparable to our $x = 0.16$ sample, which has no hump. Once this assignment is made, there is better agreement between the FS area and effective mass (Table 1). Overall, this result suggests that the small hole pocket shrinks as $x_c$ is approached from the n-type side.

Furthermore, we observe oscillations in samples which have negative MR out to 80 T, which has not been previously reported. The work from Helm et al. and Breznay et al. is predominantly at fields where the MR shows conventional, positive behaviour. This may be the reason that we observe a reduction in the oscillation amplitude below 10 K and other groups have not. A fit to the low-T regime (not shown), where the reduction occurs, results in a decrease in the Dingle lifetime ($\tau_D$) or equivalently by an increase of the Dingle temperature through the relation $T_D = \hbar/(2\pi k_B \tau_D)^{-1}$. The sudden occurrence below 10 K is attributed to a negative isotropic spin scattering contribution to the MR that occurs for $x < x_c$ and is enhanced as the doping approaches $x_c$ [26]. This spin contribution is consistent with scattering off critical fluctuations and should become stronger at lower temperatures.

**Table 1.** 80 T global fit results, using Eq. 1, for the samples presented in this work. Previous results on electron-doped cuprates are included for comparison. $F$ is the frequency of the small hole pocket in the reconstructed Fermi surface, in units of Tesla; $m^*/m_e$ is the effective mass, in units of free electron mass; $T_D$ is the Dingle temperature, in units of Kelvin; and $T^{sf}_D$ is the Dingle temperature in the regime where the oscillation amplitude is suppressed. The change in the Dingle temperature corresponds to ~ 10-20% change in the mobility for those samples.

| Sample | $F$ (T) | $m^*/m_e$ | $T_D$ (K) | $T^{sf}_D$ (K) |
|---|---|---|---|---|
| LCCO 0.11 | 305 | 0.8 | 18 | 22 |
| PCCO 0.15 | 294 | 1.0 | 14 | ---- |
| PCCO 0.16 | 267 | 0.4 | 35 | 39 |
| PCCO 0.18 | ---- | ---- | ---- | ---- |
| [25]PCO | 310 | 1.3 | 76 | ---- |
| [21]PCCO 0.14 | 255 | 0.43 | 44 | ---- |
| [18]NCCO 0.145 | 295 | 1.3 | 12.3 | ---- |
| [18]NCCO 0.15 | 292 | 1.05 | 13.5 | ---- |
| [18]NCCO 0.16 | 290 | 0.92 | 13.5 | ---- |
| [18]NCCO 0.17 | 246 | 0.88 | 13.8 | ---- |

One outstanding difference between the arguably similar PCCO samples, however, is the negative high-field MR in our sample and attendant low-temperature deviation from LK behaviour. As mentioned earlier, negative MR has been previously tied to spin scattering [26], but the sample variation is not understood. The electrical scattering rate, Hall effect, superconducting transition width, and other sample quality metrics are typical in these samples, making defects or other chemical origins difficult to blame for any differences in MR. Also, the measurement of suppressed quantum oscillation amplitude is robust and observed in different cryostats and magnets in the same PCCO sample. Given that the optimally-doped LCCO sample also exhibits similar effects, identifying the mechanism responsible for negative high-field MR appears to have important ramifications to the understanding of the FS evolution in electron-doped cuprates.

We have attributed AFM ordering at $x_c$ as responsible for the FSR and the ~300 T frequency oscillations, but the AFM correlation length, $\xi_{AFM}$ ~ 10 a, near the critical doping is quite small [3]. It is argued by Helm *et al.* [18] that the minimum correlation length would need to be ~ 18 nm (~ 45 a) in order for the charge carriers to see the ordering potential. Recent results on hole doped $YBa_2Cu_3O_{6+\delta}$ (YBCO) have demonstrated a charge density wave (CDW) ordering on the underdoped side of the phase diagram beginning at a doping p ~ 0.09 and ending at p ~ 0.16 [29, 30]. At the onset of CDW ordering (p ~ 0.09), the correlation length is $\xi_{CDW}$ ~ 10 a. This ordering is responsible for a FSR, and is supported by quantum oscillation experiments [31] and a change in the sign of the Hall coefficient [32] even though the YBCO mean free path from oscillation measurements (~ 20 a) is larger than the zero field $\xi_{CDW}$. At high fields where the oscillations are observed, the $\xi_{CDW}$ is enhanced [33] and becomes larger than the mean free path. This is a similar scenario to what we observe in the electron doped cuprates for $x \sim x_c$, i.e., short-range AFM order accompanied by a sign reversal of the Hall coefficient and a small FS measured from SdH oscillations. It is an open question, though, as to whether or not the AFM correlations are enhanced in an applied magnetic field [1] as is seen for the CDW correlations in YBCO. Without observing SdH oscillations for $x > x_c$, it is difficult to determine the precise critical doping for the FSR.

## Conclusion

To summarize, we have measured Shubnikov-de Haas oscillations on thin films of optimally doped PCCO and LCCO, as well as on overdoped PCCO, in magnetic fields up to 80 T. The oscillation frequency (~ 300 T) and effective mass (~ 1 $m_e$) of the optimally doped compounds are comparable to what has been reported for NCCO single crystals, and indicates a universal ground state electronic structure in the electron-doped cuprate family. For slightly overdoped PCCO ($x = 0.16$), we find a deviation from LK behaviour at lower temperatures, where the magnetoresistance is linear and negative up to the highest fields, indicating a reduction in the mobility most likely arising from increased/additional spin fluctuation scattering.

## Acknowledgments

This research was supported in part by the NSF under DMR-1410665 and DMR-1708334, and by NIST/Department of Commerce award 70NANB12H238. Pulsed-magnetic field measurements were partially supported by BES-DOE "Science at 100 T". The National High Magnetic Field Laboratory is supported by the National Science Foundation Cooperative Agreement No. DMR-1157490, DMR-1644779, the state of Florida, and the U.S. Department of Energy.